\documentclass{article}[11pt]
\usepackage{amsfonts}
\usepackage{CJK,amsmath,indentfirst}
\usepackage{graphics}
\usepackage{graphicx}
\begin{document}
%\begin{CJK*}{GBK}{song}

\title{\bf Residual symmetries and B\"acklund transformations}

\author{\footnotesize S. Y. Lou$^{1,2}$\\
\footnotesize $^{1}$\it Shanghai Key Laboratory of Trustworthy Computing, East China Normal University, Shanghai 200062, China\\
\footnotesize $^{2}$\it Department of Physics, Ningbo University, Ningbo, 315211, China}
\date{}
\maketitle
\parindent=0pt
\textbf{Abstract:} It is proved that for a given truncated Painlev\'e expansion of an arbitrary nonlinear Painlev\'e integrable system, the residue with respect to the singularity manifold is a nonlocal symmetry. The residual symmetries can be localized to Lie point symmetries after introducing suitable prolonged systems. The finite transformations of the residual symmetries are equivalent to the second type of Darboux-B\"acklund transformations. The once B\"acklund transformations related to the residual symmetries are same for many integrable systems including the Korteweg-de Vries, Kadomtsev-Petviashvili, Boussinesq, Sawada-Kortera and Kaup-Kupershmidt equations. For the Korteweg-de Vries equation, the $n^{th}$ Darboux transformations can also be obtained from the Lie point symmetry approach via the localization of the residual symmetries.

\vskip.4in
\renewcommand{\thesection}{\arabic{section}}
\parindent=20pt

\section{Introduction}
To understand essentially nonlinear world, scientists have introduced various idea models, integrable systems, such as the Korteweg-de Vries (KdV) equation \cite{KdV}, the nonlinear Schr\"odinger equation (NLS) \cite{NLS}, the sine-Gordon (SG) equation \cite{SG}, the Boussinesq equation \cite{Bq}, the Kadomtzev-Pedviashvili (KP) \cite{KP} equation and so on.

To study the integrability of nonlinear systems, the Painlev\'e analysis is one of the best approaches. The essential idea of the Painlev\'e analysis is to study the analytic property, however, an important aspect, the residue with respect to the singular manifold, is ignored for a longtime. Recently, it is found that the residue of the truncated Painlev\'e expansion for the bosonized supersymmetric KdV equation is a nonlocal symmetry, we call it as the residual symmetry\cite{skdv}.

In this paper, we conclude that for any Painlev\'e integrable systems, the residues of the truncated Painlev\'e expansions are all nonlocal symmetries of the original system. Another interesting common property is that for many integrable systems, there are a common residual symmetry.

We also found that some types of nonlocal symmetries can also be localized to Lie point symmetries by introducing suitable prolonged systems \cite{Point}. Here, we are consider the localizations of the residual symmetries and to find related finite transformations.

In section 2 of this paper, we first prove the existence of the residual symmetries for the Painlev\'e integrable systems and then to give out some concrete examples of the well known nonlinear systems. A common B\"acklund transformation theorem for some different nonlinear systems is also given. In section 3, taking the KdV equation as a simple example, we demonstrate that arbitrary numbers of residual symmetries can be localized to find finite transformations which is equivalent to the second type of multiple Darboux transformations. The last section is a summary and discussion.

\section{Residual symmetries of Painlev\'e integrable systems}
It is known \cite{PP} that for various integrable systems, there are possible variants which possesses the Painlev\'e property, i.e., they are Painlev\'e integrable. For the
Painlev\'e integrable systems, say, for a single component derivative polynomial system,
\begin{equation}
P(u)=0,\label{Pu}
\end{equation}
there exist a truncated Painlev\'e expansion
\begin{equation}
u=\sum_{i=0}^\alpha u_i\phi^{i-\alpha},\label{Pua}
\end{equation}
for positive integer $\alpha$.

A symmetry of \eqref{Pu} is defined as a solution of its linearized equation
\begin{equation}
P'(u)\sigma=\lim_{\epsilon=0}
\frac{\mbox{d}P(u+\epsilon\sigma)}{\mbox{d}\epsilon} =0.\label{Pus}
\end{equation}

Substituting the truncated Painlev\'e expansion \eqref{Pua} into the original nonlinear system \eqref{Pu} and vanishing the coefficients of different powers of $\phi$ we have the following theorem: \\
\bf Theorem 1. \em Residual symmetry theorem. \rm
The residue of the truncated Painlev\'e expansion \eqref{Pua} with the singular manifold $\phi$, i.e., $u_{\alpha-1}$, is a symmetry of \eqref{Pu} with the solution $u_\alpha$. \\
\em Proof. \rm Substituting \eqref{Pua} into \eqref{Pu},
we have
\begin{eqnarray}\label{Pue}
P(u)&=&P\left(\sum_{i=0}^\alpha u_i\phi^{i-\alpha}\right)\nonumber\\
&=&P(u_\alpha)+\left[P'(u_\alpha)
u_{\alpha-1}\right]\frac1\phi+O\left(\frac1{\phi^2}\right)
=0.
\end{eqnarray}
Vanishing the coefficients of $\phi^0$ and $\phi^{-1}$ respectively, we have
\begin{eqnarray}\label{Pue0}
P(u_\alpha)=0
\end{eqnarray}
and
\begin{eqnarray}\label{Pue1}
P'(u_\alpha)
u_{\alpha-1}=0.
\end{eqnarray}
Comparing \eqref{Pue0} and \eqref{Pue1} with \eqref{Pu} and \eqref{Pus} respectively, we know that $u_\alpha$ is a solution of the original model and the residue
\begin{eqnarray}\label{res}
\sigma^{u_\alpha}=u_{\alpha-1}
\end{eqnarray}
is a symmetry of \eqref{Pu} with respect to the solution $u_{\alpha}$. The theorem 1 is proved.

It is also known that for Painlev\'e integrable systems \eqref{Pu},
\begin{eqnarray}\label{up}
u_{\alpha}=u_{\alpha}(\phi)
\end{eqnarray}
changes the original nonlinear system \eqref{Pu} to its
Schwarzian form
\begin{eqnarray}\label{Pus}
P_s(\phi)=0.
\end{eqnarray}
The Schwarzian form \eqref{Pus} is invariant under the M\"obious transformation
\begin{eqnarray}\label{MT}
\phi\rightarrow \frac{a\phi+b}{c\phi+d},\ ac\neq bd.
\end{eqnarray}
which means \eqref{Pus} possesses three symmetries, $\sigma^{\phi}=a_1,\ \sigma^{\phi}=b_1\phi$ and
\begin{eqnarray}\label{f2}
\sigma^{\phi}=c_1\phi^2
\end{eqnarray}
with arbitrary constants $a_1,\ b_1$ and $c_1$. From our knowledge, the residual symmetry \eqref{res} is just related to the M\"obious transformation symmetry \eqref{f2} by the linearized equation of \eqref{up}, i.e.,
\begin{eqnarray}\label{upf}
\sigma^{u_{\alpha}}=(u_{\alpha})'_{\phi}\sigma^{\phi}
\end{eqnarray}
with \eqref{res} and \eqref{f2} while the constant $c_1$ are different for different models.

In many cases, the detailed forms of the residual symmetries can be simply fixed by the dimensional analysis. For instance, if the nonlinear system \eqref{Pu} possesses the quasilinear evolution form
\begin{eqnarray}\label{quasi}
u_t=a u_{x^b}+P_1(u,u_x,\ \cdots u_{x^c}),
\end{eqnarray}
for constants $a,\ b,$ and $c<b$, and the dimensional degree, $\delta$, of $1/u$ is smaller than or equal to $b$ (assuming the dimensional degree of $x$ is 1), then the residual symmetry of \eqref{quasi} is just
\begin{eqnarray}\label{res1}
\sigma^u=c_0\phi_{x^\delta},
\end{eqnarray}
where $c_0$ is a constant dependent on different models, hereafter we drop out the indices $\alpha$ without loss of generality.

Here are some well-known concrete examples of \eqref{quasi},\\
the KdV equation
\begin{equation}
u_t+u_{xxx}-6uu_x=0,\label{KdV}
\end{equation}
the SK equation
\begin{equation}
u_t=u_{xxxxx}+45u^2u_x+15(uu_{xx})_x=0,\label{SK}
\end{equation}
the KK equation
\begin{equation}
u_t=u_{xxxxx}+\frac{45}4uu_{xxx}+\frac{75}4u_xu_{xx}+\frac{15}2u^2u_x=0,\label{KK}
\end{equation}
and the fifth order KdV equation
\begin{equation}
u_t=u_{xxxxx}+30u^3u_x+20u_xu_{xx}+10uu_{xxx}.\label{KdV5}
\end{equation}
It is clear that for all these models the dimensional degree of $1/u$ is 2, thus the residual symmetries of all these models are
\begin{eqnarray}\label{res2}
\sigma^u=-2\phi_{xx}
\end{eqnarray}
after dropping out possible trivial constant multiplications for different models.

In fact, many other types of important nonlinear systems including the Boussinesq equation
\begin{equation}
u_{tt}+u_{xxxx}+6(uu_x)_x=0,\label{Bq}
\end{equation}
and
the KP equation
\begin{equation}
(u_t+u_{xxx}+6uu_x)_x+su_{yy}=0,\label{KP}
\end{equation}
possess the residual symmetry \eqref{res2}.

After some simple calculations, we know that the constant $c_1$ in \eqref{f2} is
\begin{equation}\label{c1}
c_1=1, \qquad (\delta=2)
\end{equation}
for all the models listed in \eqref{KdV}--\eqref{KP}.

It is known \cite{Point} that for some kinds of nonlocal symmetries, the localization procedure can be used such that the nonlocal symmetries of the original nonlinear system become local ones for a suitable prolonged system. It is interesting that if the residual symmetry is given by
\eqref{res1} and related to the M\"obious transformation symmetry \eqref{f2}, then we have the following B\"acklund symmetry theorem:\\
\bf Theorem 2. \em B\"acklund transformation theorem. \rm If the nonlinear system \eqref{Pu} possesses a Schwarzian form and the residual symmetry \eqref{res1} is related to the M\"obious transformation symmetry \eqref{f2}, then the model possesses the B\"acklund transformation
\begin{eqnarray}\label{BT1}
U(\epsilon)&=&u-\frac{c_0}{c_1}\left(\ln (c_1\epsilon f-1)\right)_{x^{\delta}},\\ \label{BT2}
F(\epsilon)&=&\frac{f}{1-\epsilon c_1 f},
\end{eqnarray}
where $\{u,\ f\}$ and $\{U(\epsilon),\ F(\epsilon)\}$ are all solutions of the nonlinear system \eqref{Pu} and its Schwarzian form. \\
\em Proof. \rm To prove the B\"acklund transformation theorem 2 is simply to solve the initial value problem:
\begin{eqnarray}\label{IC1}
\frac{\mbox{d} U(\epsilon)}{\mbox{d}\epsilon}&=&
c_0\left(F(\epsilon)\right)_{x^{\delta}},\quad U(0)=u,\\
\frac{\mbox{d} F(\epsilon)}{\mbox{d}\epsilon}&=&c_1F(\epsilon)^2,\quad F(0)=f. \label{IC2}
\end{eqnarray}
It is quite trivial to see that the solution of \eqref{IC2} is just \eqref{BT2} and substituting \eqref{BT2} into \eqref{IC1} yields \eqref{BT1} immediately. The theorem 2 is proved.

Especially, according to \eqref{res2} and \eqref{c1}, we know that the KdV, SK, KK, fifth order KdV, Boussinesq and KP equations possess a common B\"acklund transformation theorem 2 with $c_0=-2,\ c_1=1$ and $\delta=2$.

\section{B\"acklund transformations of KdV equation related to multiple residual symmetries}

For the KdV equation \eqref{KdV}, the transformation
\begin{eqnarray}\label{TK}
u=-\frac12\frac{\phi_{xxx}}{\phi_x}
+\frac14\frac{\phi_{xx}^2}{\phi_x^2}-\lambda. 
\end{eqnarray}
changes it to its Schwarzian form
\begin{eqnarray}\label{sKdV}
S+C+\lambda=0,\ C\equiv \frac{\phi_t}{\phi_x},\ S\equiv \frac{\phi_{xxx}}{\phi_x}-\frac32\frac{\phi_{xx}^2}{\phi_x^2}.
\end{eqnarray}
The residual symmetry of the KdV equation becomes a local Lie point symmetry
\begin{subequations}\label{Res}
\begin{equation}
\sigma^u=-2\phi_{xx},\label{Resu}
\end{equation}
\begin{eqnarray}
\sigma^\phi=\phi^2,\label{Resf}
\end{eqnarray}
\begin{equation}\label{Resg}
\sigma^g=2g\phi,
\end{equation}
\begin{equation}\label{Resh}
\sigma^h=2(h\phi+g^2),
\end{equation}
\end{subequations}
for the prolonged system
\begin{subequations}\label{Pr}
\begin{equation}
u_t+u_{xxx}+6uu_x=0,
\end{equation}
\begin{eqnarray}
u=-\frac12\frac{\phi_{xxx}}{\phi_x}
+\frac14\frac{\phi_{xx}^2}{\phi_x^2}-\lambda,
\end{eqnarray}
\begin{equation}\label{rg}
g=\phi_{x},
\end{equation}
\begin{equation}\label{rh}
h=g_{x}.
\end{equation}
\end{subequations}
The finite transformation of the Lie point symmetry \eqref{Res} is just the theorem 2 of the last section.

Because the symmetry equation of a nonlinear system, say, \eqref{Pus} for \eqref{Pu}, is linear and the Schwarzian form of the original nonlinear system possesses infinitely many solutions, we get infinitely many residual symmetries $\phi_{i,x^{\delta}},\ i=1,\ 2,\ \cdots$. For the KdV equation \eqref{KdV}, the infinitely many residual symmetries can be written as
\begin{equation}
\sigma_n^u=-2\sum_{i=1}^nc_i\phi_{i,xx},\label{KdVus}
\end{equation}
for arbitrary $n$, where $\phi_i,\ i=1,\ \cdots, n$ are all different solutions of the Schwarzian KdV equation \eqref{sKdV}, say for different $\lambda=\lambda_i, \ \lambda_i\neq \lambda_j,\ \forall i\neq j$.

Similar to the $n=1$ case, to find the finite transformations of \eqref{KdVus}, we have to introduce a suitable prolonged system such that the symmetry can be localized to a Lie point symmetry. Fortunately, by finishing some simple direct calculations, it is easy to find the following localization theorem:\\
\bf Theorem 3. \em Localization theorem. \rm If $\{u,\ \phi_i,\ g_i,\ h_i,\ i=1,\ 2,\ \cdots,\ n\}$ is a solution of the enlarged system
\begin{subequations}\label{Prn}
\begin{equation}
u_t+u_{xxx}+6uu_x=0,
\end{equation}
\begin{eqnarray}\label{ru}
u=-\frac12\frac{\phi_{i,xxx}}{\phi_{i,x}}
+\frac14\frac{\phi_{i,xx}^2}{\phi_{i,x}^2}-\lambda_i,
\end{eqnarray}
\begin{equation}\label{rg}
g_i=\phi_{i,x},
\end{equation}
\begin{equation}\label{rh}
h_i=g_{i,x},\ i=1,\ \cdots,\ n,
\end{equation}
\end{subequations}
then the symmetry \eqref{KdVus} is localized to a Lie point symmetry
\begin{subequations}\label{Res1}
\begin{equation}
\sigma^u=-2\sum_{j=1}^n c_jh_j,\label{Resu1}
\end{equation}
\begin{eqnarray}
\sigma^{\phi_i}=c_i\phi_i^2+\sum_{j\neq i}^n \frac{c_j}4\frac{(g_i h_j-g_j h_i)^2}{g_i g_j (\lambda_i-\lambda_j)^2},\label{Resf1}
\end{eqnarray}
\begin{equation}\label{Resg1}
\sigma^{g_i}=2c_ig_i\phi_i+\sum_{j\neq i}^n {c_j}\frac{g_i h_j-g_j h_i}{\lambda_i-\lambda_j},
\end{equation}
\begin{equation}\label{Resh1}
\sigma^{h_i}=2c_i(h_i\phi_i+g_i^2)+\sum_{j\neq i}^n\frac{c_j}2\left[4g_i g_j+\frac{(g_i^2 h_j^2-g_j^2 h_i^2)}{g_i g_j (\lambda_i-\lambda_j)}\right].
\end{equation}
\end{subequations}
\em Proof. \rm To prove the theorem 3, firstly, it is need to write down the linearized system of the enlarged system:
\begin{subequations}\label{Prns}
\begin{equation}
\sigma^u_t+\sigma^u_{xxx}+6(u\sigma^u)_x=0,
\end{equation}
\begin{eqnarray}\label{rus}
\sigma^u=-\frac12\frac{\sigma^{\phi_i}_{xxx}}{\phi_{i,x}}
+\frac12\frac{\sigma^{\phi_i}_x\phi_{i,xxx}}{\phi_{i,x}^2}
+\frac12\frac{\phi_{i,xx}\sigma^{\phi_i}_{xx}}{\phi_{i,x}^2}
-\frac12\frac{\phi_{i,xx}^2\sigma^{\phi_i}_{x}}{\phi_{i,x}^3},
\end{eqnarray}
\begin{equation}\label{rgs}
\sigma^{g_i}=\sigma^{\phi_i}_{x},
\end{equation}
\begin{equation}\label{rhs}
\sigma^{h_i}=\sigma^{g_i}_{x},\ i=1,\ \cdots,\ n.
\end{equation}
\end{subequations}
In theorem 3, there are $n$ independent but similar symmetries which are related to $c_k\neq 0,\ c_j=0,\ \forall j\neq k$ and $k=1,\ 2,\ \cdots,\ n$.

For any fixed $c_k\neq 0$ and $c_j=0,\ j\neq k$, from \eqref{Res} we know that
\begin{subequations}\label{Resk}
\begin{equation}
\sigma^u=-2c_kh_k,\label{Resuk}
\end{equation}
\begin{eqnarray}
\sigma^{\phi_k}=c_k\phi_k^2,\label{Resfk}
\end{eqnarray}
\begin{equation}\label{Resgk}
\sigma^{g_k}=2c_kg_k\phi_k
\end{equation}
\begin{equation}\label{Reshk}
\sigma^{h_k}=2c_k(h_k\phi_k+g_k^2).
\end{equation}
\end{subequations}
For $j\neq k$, eliminating $u$ from Eq. \eqref{ru} with $i=k$ and Eq. \eqref{ru} with $i=j$, we have
\begin{equation}\label{fik}
\phi_{j,xxx}=2(\lambda_k-\lambda_j)\phi_{j,x}
+\frac{\phi_{j,xx}^2}{2\phi_{j,x}}
+\frac{\phi_{k,xxx}\phi_{j,x}}{\phi_{k,x}}
-\frac{\phi_{k,xx}^2\phi_{j,x}}{2\phi_{k,x}^2}.
\end{equation}
Substituting \eqref{Resuk} into \eqref{rus} with $i=j$ and eliminating $\phi_{j,xxx}$ via \eqref{fik} yield
\begin{eqnarray}
\sigma^{\phi_j}=\frac{c_k(g_k h_j-h_k g_j)^2}{4g_j g_k(\lambda_k-\lambda_j)^2}. \label{Resfj}
\end{eqnarray}
Differentiating \eqref{Resfj} once and twice, respectively, with respect to $x$ immediately result in
\begin{eqnarray}
\sigma^{g_j}=c_k\frac{g_k h_j-g_j h_k}{\lambda_k-\lambda_j}, \label{Resgj}
\end{eqnarray}
and
\begin{eqnarray}
\sigma^{h_j}=\frac{c_k}2\left[4g_k g_j+\frac{g_k^2 h_j^2-g_j^2 h_k^2}{g_k g_j (\lambda_k-\lambda_j)}\right]. \label{Reshj}
\end{eqnarray}
After taking the linear combination of the above results for all $k=1,\ 2,\ \cdots,\ n$, the theorem 3 is proved.

Whence a nonlocal symmetry is localized to a Lie point symmetries, to find its finite transformation becomes a standard trick to solve its initial value problem according to the Lie's first principle \cite{Olver}. For the Lie point symmetry \eqref{Res1}, its corresponding initial value problem has the form
\begin{subequations}\label{IVn}
\begin{equation}
\frac{\mbox{d}U(\epsilon)}{\mbox{d}\epsilon}=-2\sum_{j=1}^n c_jH_j(\epsilon)_j,\label{IVnu}
\end{equation}
\begin{eqnarray}
\frac{\mbox{d}\Phi_i(\epsilon)}{\mbox{d}\epsilon}=
c_i\Phi_i(\epsilon)^2+\sum_{j\neq i}^n \frac{c_j}4\frac{(G_i(\epsilon) H_j(\epsilon)-G_j(\epsilon) H_i(\epsilon))^2}{G_i(\epsilon) G_j(\epsilon) (\lambda_i-\lambda_j)^2},\label{IVnf}
\end{eqnarray}
\begin{equation}\label{IVng}
\frac{\mbox{d}G_i(\epsilon)}{\mbox{d}\epsilon}
=2c_iG_i(\epsilon)\Phi_i(\epsilon)+\sum_{j\neq i}^n {c_j}\frac{G_i(\epsilon) H_j(\epsilon)-G_j(\epsilon) H_i(\epsilon)}{\lambda_i-\lambda_j},
\end{equation}
\begin{eqnarray}\label{IVnh}
\frac{\mbox{d}H_i(\epsilon)}{\mbox{d}\epsilon}
&=&2c_i[H_i(\epsilon)\Phi_i(\epsilon)+G_i^2(\epsilon)]
+\sum_{j\neq i}^n\frac{c_j}2\left[4G_i(\epsilon) G_j(\epsilon)+\frac{G_i^2(\epsilon) H_j^2(\epsilon)-G_j^2(\epsilon) H_i^2(\epsilon)}{G_i(\epsilon) G_j(\epsilon) (\lambda_i-\lambda_j)}\right],
\end{eqnarray}
\begin{equation}\label{IVn0}
U(0)=u,\ \Phi_i(0)=\phi_i,\ G_i(0)=g_i,\ H_i(0)=h_i,\ i=1,\ \cdots,\ n.
\end{equation}
\end{subequations}
After solving out the initial valued problem \eqref{IVn}, we get the following $n^{th}$ B\"aclund transformation theorem. \\
\bf Theorem 4. \em $n^{th}$ B\"aclund transformation theorem. \rm If $\{u,\ \phi_i,\ g_i,\ h_i,\ i=1,\ 2,\ \cdots,\ n\}$ is a solution of the prolonged KdV system \eqref{Prn}, so is $\{U(\epsilon),\ \Phi_i(\epsilon),\ G_i(\epsilon),\ H_i(\epsilon),\ i=1,\ 2,\ \cdots,\ n\}$
where
\begin{subequations}\label{nBT}
\begin{equation}
U(\epsilon)=u+2(\ln \Delta)_{xx},\label{nBTu}
\end{equation}
\begin{eqnarray}
\Phi_i(\epsilon)=-\frac{\Delta_i}{\Delta},\label{nBTf}
\end{eqnarray}
\begin{equation}\label{nBTg}
G_i(\epsilon)=\Phi_{i,x}(\epsilon),
\end{equation}
\begin{equation}\label{Reshk}
H_i(\epsilon)=\Phi_{i,xx}(\epsilon)
\end{equation}
\end{subequations}
with $\Delta$ and $\Delta_i$ are the determinants of
\begin{equation}\label{M}
M=\left(\begin{array}{cccccc}
c_1\epsilon \phi_1-1 & c_1\epsilon w_{12} & \cdots & c_1\epsilon w_{1j} & \cdots & c_1\epsilon w_{1n}\\
c_2\epsilon w_{12}& c_2\epsilon \phi_2-1 & \cdots & c_2\epsilon w_{2j} & \cdots & c_2\epsilon w_{2n}\\
\vdots & \vdots & \vdots & \vdots & \vdots & \vdots\\
c_j\epsilon w_{1j} & c_j\epsilon w_{2j} & \cdots &c_j\epsilon \phi_j-1 & \cdots & c_j\epsilon w_{jn}\\
\vdots & \vdots & \vdots & \vdots & \vdots & \vdots \\
c_n\epsilon w_{1n} & c_n\epsilon w_{2n} & \cdots & c_n\epsilon w_{jn}& \cdots & c_n\epsilon \phi_n-1
\end{array}\right),\quad w_{ij}\equiv\frac{g_ih_j-g_jh_i}
{2\sqrt{g_ig_j}(\lambda_i-\lambda_j)},
\end{equation}
\begin{equation}\label{Mi}
M_i=\left(\begin{array}{cccccccc}
c_1\epsilon \phi_1-1 & c_1\epsilon w_{12} & \cdots & c_1\epsilon w_{1,i-1} &c_1\epsilon w_{1i} &c_1\epsilon w_{1,i+1} & \cdots & c_1\epsilon w_{1n}\\
c_2\epsilon w_{12}& c_2\epsilon \phi_2-1 & \cdots & c_2\epsilon w_{2,i-1} &c_2\epsilon w_{2i} &c_2\epsilon w_{2,i+1} & \cdots & c_2\epsilon w_{2n}\\
\vdots & \vdots & \vdots & \vdots & \vdots & \vdots & \vdots & \vdots\\
c_{i-1}\epsilon w_{1,i-1} & c_{i-1}\epsilon w_{2,i-1} & \cdots &c_{i-1}\epsilon \phi_{i-1}-1 &c_{i-1}\epsilon w_{i-1,i} &c_{i-1}\epsilon w_{i-1,i+1} & \cdots & c_{i-1}\epsilon w_{i-1,n}\\
 w_{1i} &  w_{2i} & \cdots & w_{i,i-1} & \phi_i & w_{i,i+1} & \cdots &  w_{in}\\
c_{i+1}\epsilon w_{1,i+1} & c_{i+1}\epsilon w_{2,i+1} & \cdots &c_{i+1}\epsilon w_{i-1,i+1} &c_{i+1}\epsilon w_{i,i+1} &c_{i+1}\epsilon \phi_{i+1}-1 & \cdots & c_{i+1}\epsilon w_{i+1,n}\\
\vdots & \vdots & \vdots & \vdots &\vdots &\vdots & \vdots & \vdots \\
c_n\epsilon w_{1n} & c_n\epsilon w_{2n} & \cdots & c_n\epsilon w_{i-1,n}&c_n\epsilon w_{in}&c_n\epsilon w_{i+1,n}& \cdots & c_n\epsilon \phi_n-1
\end{array}\right).
\end{equation}
Because the Schwarzian KdV system \eqref{sKdV} possesses the M\"obious transformation invariance \eqref{MT} it is clearly that group parameters $\epsilon$ and $c_i,\ \forall i=1,\ \cdots,\ n$ in the theorem 4 can be simply taken as $1$ and $c_j\epsilon f_j-1$ can also be simply replaced by $f_j$ without loss of generality.

According to the theorem 4, starting from any seed solutions of the KdV equation and Schwarzian form, we can obtain infinitely many new solutions. For instance, starting from the trivial vacuum solution
\begin{equation}\label{u0}
u=0,
\end{equation}
we can find a special solution of \eqref{ru} and the Schwarzian KdV \eqref{sKdV} with $\lambda=\lambda_i$ in the form
\begin{equation}\label{pi}
\phi_i=\exp(k_ix-k_i^3t),\ \lambda_i=-\frac14 k_i^2,
\end{equation}
and then
\begin{equation}\label{swij}
w_{ij}=\frac{2\sqrt{k_ik_j}}{k_i+k_j}
\exp\left\{\frac{k_i+k_j}2[x+(k_1^2-k_1k_2+k_2^2)t]\right\}.
\end{equation}
Substituting the special solution \eqref{pi} and \eqref{swij} with \eqref{u0} into the theorem 4, we reobtain the multiple soliton solutions of the KdV equation, say,
\begin{equation}\label{2s}
\Delta=1-c_1a\exp{(k_1x-k_1^3t)}-c_2a\exp{(k_2x-k_2^3t)}
+c_1c_2a^2\frac{(k_1-k_2)^2}{(k_1+k_2)^2}
\exp{[(k_1+k_2)x-(k_1^3+k_2^3)t]}
\end{equation}
for the two-soliton solution. 

Furthermore, because of \eqref{ru}, it is easy to prove that $w_{ij}$ in \eqref{M} satisfy
\begin{equation}
w_{ij,x}=\sqrt{g_ig_j}=\psi_i\psi_j,\ \forall i,j=1,\ 2,\ \cdots,\ n,
\end{equation}
where $\psi_i,\ i=1,\ 2,\ \cdots,\ n$ are just the usual spectral function of the Lax pair
\begin{equation}
\psi_{i,xx}+(u+\lambda_i)\psi_i=0.\ i=1,\ 2,\ \cdots,\ n
\end{equation}
of the KdV equation.
Thus the theorem 4 is equivalent to the known second type of Darboux transformation \cite{DT}. In other word the theorem 4 can be restated as: \\
\bf Theorem 5. \em Equivalent DT theorem. \rm If $u$ is a solution of the KdV equation and $\psi_i,\ i=1,\ \cdots,\ n$ are spectral functions related to $u$ with spectral parameter $\lambda_i$, so is $u_n$ with
\begin{equation}
u_n=u+2[\ln (\det M)]_{xx},\ M_{ij,x}=\psi_i\psi_j. \label{DT}
\end{equation}

\section{Summary and discussions.}
In summary, it is shown that for Painlev\'e integrable systems, infinitely many nonlocal symmetries which is defined as residual symmetries can be readily read out from the residual of the truncated Painlev\'e expansions.
The residual symmetries are nonlocal for the original nonlinear system. However, the residual symmetries can be readily localized to Lie point symmetries by prolonging the original system to a larger one.

If the form of the residual symmetries are same for different nonlinear systems, a common first B\"acklund transformation can be obtained. For instance, for the KdV equation \eqref{KdV}, the fifth order KdV equation \eqref{KdV5}, the SK system \eqref{SK}, the KK equation \eqref{KK}, the Boussinesq equation \eqref{Bq}, the KP system \eqref{KP} and so on, the once B\"acklund transformation possess the same theorem 3 with $c_0=-2,\ c_1=1$ and $\delta=2$.

The explicit forms of the finite transforms for $n$ residual symmetries are obtained for the KdV equation. The result is equivalent to the second type of $n^{th}$ B\"acklund transformation \cite{DT}. Though the theorem 4 is equivalent to the known DT theorem 5, some further points should be emphasized: (1). The second type of DT and then multiple soliton solutions can be obtained via Lie point symmetry method companied by the localization approach. (2). The group parameter, $\epsilon$, is necessary to find the second type of DT via Lie point symmetry approach, however, it is only an auxiliary one because it can be absorbed by the M\"obious transformation invariance of the Schwarzian systems. And the other group parameters $c_i$ can also be absorbed. (3). The commutable theorem(s) for B\"acklund transformations and Darboux transformations becomes a  trivial fact in Lie point symmetry theory because of the   commutativity in addition algorithm. The more about the residual symmetries will be further studied in our future researches.

\section*{Acknowledgement.}
The authors are in debt to thanks the helpful discussions  with Dr. Y. Q. Li, J. C. Chen, Profs. X. B. Hu, X. Y. Tang and Y. Chen.
The work was sponsored by the National Natural Science Foundations of
China (Nos. 11175092, 11275123, 11205092 and 10905038), Shanghai Knowledge Service Platform for Trustworthy Internet of Things (No. ZF1213), Talent Fund
and K. C. Wong Magna Fund in Ningbo University.

\small{
}
\end{document}